\documentstyle[aps,prl,epsfig]{revtex}
\tightenlines

\newcommand{\be}{\begin{equation}}
\newcommand{\ee}{\end{equation}}

\newcommand{\bmath}{\begin{mathletters}}
\newcommand{\emath}{\end{mathletters}}

\begin{document}

\title{\Large{\bf Magnetic Ordering of Itinerant Systems in Modified CPA Approximation}}

\vskip0.5cm 

\author{ G. G\'{o}rski, J. Mizia and K. Kucab }

\address{Institute of Physics, University of Rzesz\'{o}w, ulica Rejtana 16A, \\
35-958 Rzesz\'{o}w, Poland\\}

\maketitle

\vskip0.5cm 


\begin{abstract}

\noindent
 We will analyze the itinerant model for ferromagnetism with both single-site and two-site electron correlations. We will include band degeneration into the model. This will allow 
us to consider the on-site exchange interactions in the Hamiltonian. The modified Hartree-Fock approximation for the two-site interactions will be used. This approximation will 
give the relative one spin band broadening with respect to the other, in addition to the shift in position of majority and minority spin bands. 
Next, we will use the coherent potential approximation technique (CPA) with two self-energies. One will describe the on-site correlation and the second one the inter-site 
correlation. We will separate from these self-energies, the linear terms arising from Hartree-Fock approximation and the higher order terms. The on-site linear term will contribute 
only to the effective molecular field. The inter-site linear term will contribute to the effective molecular field and to the inter-site correlation factor. 
The Green function technique and CPA decoupling will allow for the change in the shape of spin bands, which has been described by the correlation factors and which will 
decrease the kinetic energy of the system. The gain in kinetic energy due to the on-site and inter-site correlation factors will drive the ferromagnetism and significantly reduce the 
effective Hartree field necessary to create this ordering.

\end{abstract}

\vskip1.5cm


\noindent {\Large {\bf 1. Introduction}}

\vskip0.5cm 

The basic model for magnetic ordering of itinerant electrons in solids is the Hubbard model \cite{1}. In the mean-field approximation, the Hubbard model leads to the well-known 
Stoner model for magnetism \cite{2,3}. The Coulomb constant $U$ coming out of the Stoner condition for creating ferromagnetism is large, i.e. of the order of bandwidth. On the 
one hand it can be justified by the existing strong Coulomb interaction, but on the other hand for such a strong interaction, one can not use mean-field approximation.
This has prompted attempts to treat the problem within the higher order perturbation theory. From the many approaches considered by others, we mention only Hubbard I and 
conventional CPA \cite{4}, which like many other approaches have failed to bring any type of ferromagnetic ordering \cite{5}. They did not produce the spin-dependent band 
shift necessary for a ferromagnetic ordering which is why new versions of the conventional CPA are still being created.  
One of these new attempts is a self-consistent moment method called "spectral density approach" (SDA) \cite{6}. The magnetic phase diagrams calculated by this method are 
more realistic and the values of the Curie temperatures also make sense. The main advantage of SDA is obtaining a spin-dependent band shift. The main shortcoming of this 
method is the lack of quaziparticle damping. Therefore, Nolting and co-workers have proposed a combination of SDA and CPA called "modified alloy analogy" (MAA) \cite{7}.
Nonetheless, in our knowledge there is no reliable solution of the magnetic ordering problem in the Hubbard model, and still to this day the problem has remained unsolved. 
More recently, the new dynamical mean-field theory (DMFT) \cite{8} has been developed for a direct computational simulation of systems with correlated electrons on a crystal 
lattice. This method has an exact solution in the non-trivial limit of an infinite coordination number \cite{9}. The results have been obtained by using direct quantum Monte-Carlo 
(QMC) simulation and the mean-field Green function theory. Use of this DMFT method, has introduced a significant progress in the theory of ferromagnetism. The results (see e.g. 
\cite{10,11}) show the existence of ferromagnetism but at much lower temperatures than those coming from the Hartree-Fock approximation. Such results would remove the 
problem known as a "magnetic paradox", i.e. the Curie temperature coming from interaction constant $U$-fitted to obtain the proper magnetic moment at $T=0K$- as being much 
too high.
In this paper we will use a more conventional approach of many body theory. We included in the CPA method both the on-site and inter-site correlations. We will show, despite 
the use of a traditional approach, that qualitatively new results can be calculated which can bring the constant of the mean field creating magnetism to almost zero. 
Extension of the Hubbard model is the model which, in addition to the on-site Coulomb repulsion $U$ includes the on-site exchange interaction $J_{in}=(i\lambda,i\nu\mid 1/r\mid 
i\nu,i\lambda)$ (for  $\lambda  \ne \nu$, where $\lambda$, $\nu$ are the sub-band indices) and the inter-site exchange interaction $J=(i\lambda,j\nu\mid 1/r\mid j\nu,i\lambda)$ 
\cite{12}. To obtain a magnetic state, many of the previous papers on metallic ferromagnetism have utilized the on-site exchange interaction (see e.g. \cite{12,13}). Hirsch  
\cite{14,15,16,17,18} and others (see e.g. \cite{19,20,21}) included the nearest-neighbor inter-site interactions without the band degeneracy; $(i,j\mid 1/r\mid k,l)$, where $(k,l)=(i,j)$ 
and $i$, $j$ are the nearest-neighbor lattice sites. Their solutions of the 3 dimensional Hubbard model did not yield the exchange field necessary for ordering which would have 
been small enough to be justified physically. 
In our model, we introduce into Hamiltonian the on-site Coulomb repulsion $U$, single-site two sub-band interactions $\left( {i\lambda ,j\nu |1/r|k\vartheta ,l\varpi } \right)$ 
($i=j=k=l$ and for the sub-band indices we have the restriction $(\varpi ,\vartheta ) = (\lambda ,\nu )$ and $\lambda  \ne \nu$), and also the inter-site nearest-neighbor interactions 
$(i,j\mid 1/r\mid k,l)$ (for $(k,l)=(i,j)$ and $i \ne j$).
Analyzing the inter-site correlations, we deal with the probabilities defined for the product of four operators (the two operator product is replaced by its stochastic value). We 
have interpreted stochastic value of the two operators product as the product of probabilities for electron transfer between two atoms in the presence of another electron. This 
interpretation makes the connection between Hirsch's \cite{14} average bond occupation for spin $\sigma$; $I_\sigma$, and the standard probabilities used in the CPA method 
\cite{4}. This new link allows us to expand the model to the more realistic cases of the weak inter-site interaction being accompanied by the single-site Coulomb interaction of any 
strength. The spin band narrowing (broadening) is arising from the two-site interactions (see \cite{14,22}) and also from the on-site interaction (see \cite{5,23}). On the basis of 
general understanding, how the correlation affects the density of states (DOS) (see e.g. Fukuyama \cite{23}), we can interpret the part of the band broadening which is coming 
from the two-site interactions, as being expressed by the inter-site correlation factor $K_{ij}$, even in the lowest-first order approximation.

The paper is organized as follows. In Section 2, we have put forward the model Hamiltonian and developed the CPA formalism to treat the on-site and inter-site Coulomb 
correlation at the same time. In Section 3, we have set up the model for ferromagnetism, which includes the on-site Coulomb correlation, inter-site correlation and the assisted 
hopping correlation. Numerical examples are presented in Section 4 based on the semi-elliptic DOS for the weak inter-site interactions, in the presence of on-site Coulomb 
interaction of any strength, with and without hopping correlation. On the basis of these results, the conclusions regarding the appearance of magnetic ordering with growing 
occupation of the band are drawn in Section 5.

\vskip1cm 

\noindent {\Large {\bf 2. Hamiltonian and the Coherent Potential Approximation}} 

\vskip0.5cm 

The Hamiltonian for one degenerate band can be written in the form given by Hubbard \cite{24} 
\be 
H =  - \sum\limits_{\scriptstyle ij,\lambda \nu  \hfill \atop 
  \scriptstyle \,\,\,\,\sigma  \hfill} {t_{ij} \left( {c_{i\lambda \sigma }^ +  c_{j\nu \sigma }  + h.c} \right)}  - \mu _0 \sum\limits_{i,\lambda ,\sigma } {\hat n_{i\lambda \sigma } }  + 
\sum\limits_{\scriptstyle \,\,\,\,ijkl, \hfill \atop 
  {\scriptstyle \lambda \nu \varpi \vartheta  \hfill \atop 
  \scriptstyle \,\, \sigma ,\sigma ' \hfill}} {\left( {i\lambda ,j\nu |1/r|k\vartheta ,l\varpi } \right)c_{i\lambda \sigma }^ +  c_{j\nu \sigma '}^ +  c_{l\varpi \sigma '} c_{k\vartheta \sigma 
} } ,
\label{1}
\ee 

\noindent
where $t_{ij}$ is the nearest neighbors hopping integral, $\mu _0$ is the chemical potential, $c_{i\lambda \sigma }^ +  \left( {c_{i\lambda \sigma } } \right)$ creates (destroys) an 
electron of spin $\sigma$   in a Wannier orbital $\lambda$ on the $i$-th lattice site, the indices $\lambda ,\nu ,\varpi ,\vartheta$ numerate the sub-bands in the degenerated single 
band. 
Taking into account in Hamiltonian (\ref{1}) only single-site $i=j=k=l$ and two-site interactions $((k,l)=(i,j))$, as well as single sub-band $(\lambda=\nu=\varpi=\vartheta)$ and two 
sub-band interactions $((\varpi ,\vartheta )=(\lambda ,\nu ))$ we have obtained and retained the following matrix elements; \\
\begin{itemize}
\bmath
 \item single-site, single sub-band interaction
\be
U_0  = \left( {i\lambda ,i\lambda |1/r|i\lambda ,i\lambda } \right) \label{2a},
\ee
 \item single-site (subscript "in"), two sub-band interactions (for $\lambda  \ne \nu $ )
\be
 V_{in}  = \left( {i\lambda ,i\nu |1/r|i\lambda ,i\nu } \right),\,\,\,\, J_{in}  = \left( {i\lambda ,i\nu |1/r|i\nu ,i\lambda } \right),\,\,\,\, J'_{in}  = \left( {i\lambda ,i\lambda |1/r|i\nu ,i\nu } 
\right) \label{2b},
\ee
 \item two-site interactions
\[
V_0  = \left( {i\lambda ,j\nu |1/r|i\lambda ,j\nu } \right),\,\,\,\,\, J_0  = \left( {i\lambda ,j\nu |1/r|j\nu ,i\lambda } \right),
\]
\be
 \;\;\;\;\;\;\;\;J'_0  = \left( {i\lambda ,i\lambda |1/r|j\nu ,j\nu } \right),\,\,\,\,\, \Delta t_0  = \left( {i\lambda ,i\lambda |1/r|j\nu ,i\lambda } \right) \label{2c}.
\ee
\emath
 \end{itemize}
 
 The mean-field approximation was applied to intra-atomic weak interactions $J_{in}$, $J'_{in}$, $V_{in}$ and to the assisted hopping interaction, $\Delta t_0$. After this 
operation, this part of the interaction together with the kinetic energy of Hamiltonian (\ref{1}) took on the following form
\be
H_0  =  - \sum\limits_{\scriptstyle ij \hfill \atop 
  \scriptstyle \sigma  \hfill} {t_{ij}^\sigma  \left( {c_{i\sigma }^ +  c_{j\sigma }  + h.c.} \right)}  - \sum\limits_{i\sigma } {\mu _0 } \hat n_{i\sigma }  + \sum\limits_{i\sigma } 
{M^\sigma  \hat n_{i\sigma } }, 
\label{3}
\ee
 where the molecular (mean) field is given by
 \be
M^\sigma   = In^{ - \sigma }  + 2z\Delta tI_{ - \sigma } (m),
\label{4}
\ee
 with $I$ being the effective on-site exchange interaction constant
 \be
I = \left( {d - 1} \right)\left( {J_{in}  + J'_{in}  + V_{in} } \right),
\label{5}
\ee
 where $d$ is the number of sub-bands, and $I_{ - \sigma } \left( m \right)$ the average bond occupation for spin $-\sigma$ and magnetization $m$
 \be
I_{ - \sigma } (m) = \left\langle {c_{i - \sigma }^ +  c_{j - \sigma } } \right\rangle,
\label{6} 
\ee  
 where $z$ is the number of nearest-neighbors. The spin-dependent hopping integral $t_{ij}^\sigma$ is expressed by
 \be
t_{ij}^\sigma   = t_{ij}  - \Delta t\left( {n_{i - \sigma }  + n_{j - \sigma } } \right), \;\;\;\; \Delta t = d \cdot \Delta t_0.
\label{7}
 \ee
 The approximation given by Eq. (\ref{7}), changing the bandwidth, was proposed by Micnas \cite{22} and Hirsch \cite{14,15}.
 
After the Fourier transform we obtain
\be
H_0  = \sum\limits_{k\sigma } {(\varepsilon _k^\sigma   - \mu )\hat n_{k\sigma } },
\label{8}
\ee
where
\be
\varepsilon _k^\sigma   = \varepsilon _k \left( {1 - 2\frac{{\Delta t}}{t}n^{ - \sigma } } \right) - \sigma \frac{{M_m^\sigma  }}{2}m,
\label{9}
\ee
\be
\varepsilon _k =- ts_k, \;\;\;\;s_k  = \sum\limits_{ < i,j > } {e^{ik\left( {R_i  - R_j } \right)} },\;\;\;\;  \mu  = \mu _0  - \frac{{M_0 }}{2}n
\label{10}
\ee
 and $t = t_{ij}$ is the nearest-neighbor hopping integral.
 
The modified molecular field is 
\be
M^\sigma   = \frac{{M_0 }}{2}n - \sigma \frac{{M_m^\sigma  }}{2}m,
\label{11}
\ee
where $M_{0\left( m \right)}$ have the following forms:
\be
M_0  = I + \frac{{2z\Delta t}}{n}I_0,
\label{12}
\ee
 \be
M_m^\sigma   = I - \sigma \frac{{2z\Delta t}}{m}\left( {I_{ - \sigma } (m) - I_0 } \right),
\label{13}
\ee
 where
 \[
I_0  = \mathop {\lim }\limits_{m \to 0} I_{ \pm \sigma } (m).
\]

We will leave now, for the time being, the single-site Coulomb repulsion and only consider, in the Hamiltonian (\ref{1}), the inter-site terms and small intra-atomic terms; $J_{in}$, 
$J'_{in}$, $V_{in}$, treated in Hartree-Fock approximation. The following form was arrived at
\be
H = \sum\limits_{k\sigma } {(\varepsilon _k^\sigma   - \mu )\hat n_{k\sigma } }  + \sum\limits_{ < ij > \sigma } {\left[ {Jc_{j - \sigma }^ +  c_{i - \sigma }  + J'c_{i - \sigma }^ +  c_{j 
- \sigma }  + Jc_{j\sigma }^ +  c_{i\sigma }  - Vc_{j\sigma }^ +  c_{i\sigma } } \right]c_{i\sigma }^ +  c_{j\sigma } },
\label{14} 
\ee
where the new inter-site constants $J$, $J'$, $V$ for the degenerate band are given by $dJ_0$, $dJ'_0$, $dV_0$ respectively, where $d$ is the number of degenerated orbitals in the 
band. The physical meaning of these interactions is the following; $J$ is the exchange interaction, $J'$ the pair-hopping interaction, $V$ the density-density interaction.

The CPA idea was applied to the Hamiltonian (\ref{14}) and the stochastic value replaced each operator product in the square bracket. For example, the operators product; 
$c_{j\sigma }^ +  c_{i\sigma }$ was replaced by the stochastic value $\overline {c_{j\sigma }^ +  c_{i\sigma } }$. It took the values 1or 0. The probability of value 1 is the 
probability of the electron with spin $\sigma$ hopping from the $i$ to the $j$ lattice site which is given by the product of probabilities that there is an electron with spin $\sigma$ 
on the $i$ site and that the $j$ site is empty; $n_i^\sigma  (1 - n_j^\sigma )$. This probability is called $P_1^\sigma$ below.

Introducing the inter-site self-energy $\Sigma _{1,2}^\sigma$ it can be written that
\be
H = \sum\limits_{k\sigma } {E_k^\sigma  \hat n_{k\sigma } }  + \sum\limits_{\scriptstyle \left\langle {i,j} \right\rangle  \hfill \atop 
  \scriptstyle   \sigma  \hfill} {\left( {\varepsilon _i  - \Sigma _{1,2}^\sigma  } \right)c_{i\sigma }^ +  c_{j\sigma } },
\label{15}
\ee
where the effective dispersion relation (see Eqs (\ref{9}) and (\ref{10})) is given by
\be
E_k^\sigma   = \varepsilon _k^\sigma   - \mu  + s_k \Sigma _{1,2}^\sigma   = \varepsilon _k \left( {1 - \frac{{\Sigma _{1,2}^\sigma  }}{t} - 2\frac{{\Delta t}}{t}n^{ - \sigma } } 
\right) - \sigma \frac{{M_m^\sigma   + zJ}}{2}m - \mu.
\label{16} 
\ee
According to what was said above and from Eq. (\ref{14}) the stochastic potential $\varepsilon _i$ is equal to
\[
\varepsilon _i  = J\overline {c_{j - \sigma }^ +  c_{i - \sigma } }  + J'\overline {c_{i - \sigma }^ +  c_{j - \sigma } }  + J\overline {c_{j\sigma }^ +  c_{i\sigma } }  - V\overline 
{c_{j\sigma }^ +  c_{i\sigma } }. 
\]
 It will take on one of the following values
 \be
\varepsilon _i  = \left\{ \begin{array}{l}
 \varepsilon _1  = J - V \\ 
 \varepsilon _2  = J + J' \\ 
 \varepsilon _3  = 0 \\ 
 \end{array} \right.; 
 {\rm \; with \; probabilities; \;\;} 
\begin{array}{l}
 P_1^\sigma   = I_\sigma  (m) = n_j^\sigma  \left( {1 - n_i^\sigma  } \right) \\ 
 P_2^\sigma   = I_{ - \sigma } (m) = n_j^{ - \sigma } \left( {1 - n_i^{ - \sigma } } \right) \\ 
 P_3^\sigma   = 1 - P_1^\sigma   - P_2^\sigma   \\ 
 \end{array}.
 \label{17}
  \ee
 $P_1^\sigma$ and  $P_2^\sigma$ above are equal to the quantities $I_\sigma$, $I_{-\sigma}$ in Ref. \cite{14}, if we assume the ferromagnetic order, i.e. $n_i^\sigma=n_j^\sigma  
=n^\sigma$. Once again, e.g. the probability $P_2^\sigma$ is the probability of electron with spin $-\sigma$ hopping from the $j$ to the $i$ lattice site in the presence of another 
electron with spin $\sigma$ (hopping from the $j$ to the $i$ site) detected by the operators product $c_{i\sigma }^ +  c_{j\sigma}$, which is at the end of Eq. (\ref{14}).

With these probabilities and potentials one can write the following equation for the self-energy $\Sigma _{1,2}^\sigma$, which describes the inter-site interactions only 
(Hamiltonian (\ref{15}))
\be
\sum\limits_{i = 1}^3 {P_i^\sigma  \frac{{\varepsilon _i  - \Sigma _{1,2}^\sigma  }}{{1 - \left( {\varepsilon _i  - \Sigma _{1,2}^\sigma  } \right)G^\sigma  }}}  = 0,
\label{18}
\ee
where $P_i^\sigma$ and $\varepsilon _i$ are given by Eq. (\ref{17}) and the Slater-Koster function has the following form
\be
G^\sigma  \left( \varepsilon  \right) = \frac{1}{N}\sum\limits_k {\frac{1}{{\varepsilon  - \varepsilon _k^\sigma   + \mu  - s_k \Sigma _{1,2}^\sigma  }}}.
\label{19} 
\ee
In the first approximation, it was obtained from Eq. (\ref{18}) that
\be
\Sigma _{1,2}^\sigma   \cong \bar \varepsilon _{1,2}^\sigma   = P_1^\sigma  \varepsilon _1  + P_2^\sigma  \varepsilon _2.
\label{20}
\ee
The effective dispersion relation in the first order approximation is given on the base of Eq. (\ref{16}) and (\ref{20}) by the relation
\begin{eqnarray}
E_k^\sigma  &=& \varepsilon _k^\sigma   + s_k \bar \varepsilon _{1,2}^\sigma   - \mu  \nonumber\\ 
 &=&\varepsilon _k \left[ {1 - \frac{{2z\Delta t}}{D}n^{ - \sigma }  - \frac{{z\varepsilon _1 }}{D}n^\sigma  (1 - n^\sigma  ) - \frac{{z\varepsilon _2 }}{D}n^{ - \sigma } (1 - n^{ - 
\sigma } )} \right] - \sigma \frac{{M_m^\sigma   + zJ}}{2}m - \mu,  
\label{21}
\end{eqnarray}
which is identical to Eq. (\ref{4}) in Ref. \cite{15} with the exception of the $\Delta t$ term included here. One can see that this result is only the first-order approximation of the 
self-energy $\Sigma _{1,2}^\sigma$ in $J/D$, $J'/D$ and $V/D$, where $D=zt$ is the half-bandwidth and $z$ is the number of nearest-neighbors. Equation (\ref{18}) allows for the 
higher order calculations of the inter-site interactions through the inter-site self-energy $\Sigma _{1,2}^\sigma$.

It is worth mentioning here that in the first order approximation (with respect to the differences $\left( {\varepsilon _j  - \varepsilon _i } \right)/D$) the self-energy $\Sigma 
_{1,2}^\sigma$ in Eq. (\ref{20}) depended on the spin index $\sigma$, but as is known today (see Refs \cite{5,23,25}), the mean-field approximation tends to overestimate the 
effects of ordering. When summed up, the infinite CPA series, coming from Eq. (\ref{18}) and the self-energy dependence on spin index disappeared, i.e. $\Sigma _{1,2}^\sigma   = 
\Sigma _{1,2}^{ - \sigma }  \equiv \Sigma _{1,2}$. As a consequence, the full solution of Eq. (\ref{18}) has given the ground state which remained paramagnetic until an additional 
molecular field coming from the Hartree-Fock approximation (used for some single- and two-site interactions) was introduced into the Hamiltonian (\ref{1}). This molecular field 
brought the energy dependence on $\sigma$ through the field $\sigma \frac{{M_m^\sigma  }}{2}m$ (see Eq. (\ref{13})).

To include effect of on-site Coulomb repulsion, in addition to the two-site interactions, it was now assumed that the effective stochastic atomic energy $E_i^\sigma$ at the $i$-th 
site is given by (see Ref. \cite{23})
\be
E_i^\sigma   = \Sigma _U^\sigma   + \varepsilon _i, 
\label{22}
\ee
where $\varepsilon _i$ is given by Eq. (\ref{17}) and the probabilities of energies $E_i^\sigma$  are also given by Eq. (\ref{17}), but now the Eq. (\ref{18}) is replaced by
\be
\sum\limits_{i = 1}^3 {P_i^\sigma  \frac{{E_i^\sigma   - \Sigma _{1,2}^\sigma  }}{{1 - \left( {E_i^\sigma   - \Sigma _{1,2}^\sigma  } \right)F^\sigma  }}}  = 0,
\label{23}
\ee
with
\be
F^\sigma   = \frac{1}{N}\sum\limits_k {\frac{1}{{\varepsilon  - \varepsilon _k^\sigma   + \mu  - s_k \Sigma _{1,2}^\sigma   - \Sigma _U^\sigma  }}} 
\label{24}
\ee
and with the single-site self energy given by 
\be
\Sigma _U^\sigma   = \frac{{Un^{ - \sigma } }}{{1 - \left( {U - \Sigma _U^\sigma  } \right)F^\sigma  }},
\label{25}
\ee
where the zero energy is defined at the atomic level.

It is worth noting here that because of the additional factor $s_k$ in Eq. (\ref{24}), there was a modification to the standard method of calculating the DOS. When we used the 
relation $s_k=-\varepsilon _k/t$, and the relation (\ref{9}) for   we were be able to write

\begin{eqnarray}
F^\sigma  \left( \varepsilon  \right) & =& \frac{1}{N}\sum\limits_k {\frac{1}{{\varepsilon  - \varepsilon _k \left( {1 - 2\frac{{\Delta t}}{t}n^{ - \sigma }  - \frac{{\Sigma 
_{1,2}^\sigma  }}{t}} \right) + \sigma \frac{{M_m^\sigma   + zJ}}{2}m - \Sigma _U^\sigma   + \mu }}}   \nonumber \\  
 &=&\frac{1}{{b^\sigma  }}F_0 \left( {\frac{{\varepsilon  - \Sigma _U^\sigma   + \sigma \frac{{M_m^\sigma   + zJ}}{2}m + \mu }}{{b^\sigma  }}} \right),
\label{26}
\end{eqnarray}
where the unperturbed Slater-Koster function is given by $F_0 \left( \varepsilon  \right) = \frac{1}{N}\sum\limits_k {\frac{1}{{\varepsilon  - \varepsilon _k }}}$ and
\be
b^\sigma   = 1 - \frac{{\Sigma _{1,2}^\sigma  \left( \varepsilon  \right)}}{t} - 2\frac{{\Delta t}}{t}n^{ - \sigma }. 
\label{27}
\ee
In our derivation we made the assumption that both self-energies were $k$-independent, while they can be energy-dependent.

\vskip0.5cm 
\noindent {\Large {\bf 3.  Ferromagnetism (F)}} 
\vskip0.5cm 

At nonzero temperature the number of electrons with spin $\sigma$ is given by 
\be
n^\sigma   =  - \int\limits_{ - \infty }^\infty  {f\left( \varepsilon  \right)\frac{1}{\pi }{\mathop{\rm Im}\nolimits} F^\sigma  \left( \varepsilon  \right)d\varepsilon }, 
\label{28}
\ee
where $f(\varepsilon)$ is the Fermi function
\be
f\left( \varepsilon  \right) = \frac{1}{{1 + e^{\beta \left[ {\varepsilon  - \left( {\mu  + \sigma \frac{{M_m^\sigma   + zJ}}{2}m} \right)} \right]} }}
\label{29}
\ee
and
\be
F^\sigma  \left( \varepsilon  \right) = \frac{1}{N}\sum\limits_k {\frac{1}{{\varepsilon  - \varepsilon _k b^\sigma  (\varepsilon ) - \Sigma _U^\sigma  \left( \varepsilon  \right)}}}. 
\label{30}
\ee

The energy $\varepsilon$ in above expressions is from now on equal to $\varepsilon  + \mu  + \sigma m(M_m^\sigma   + zJ)/2$.

To obtain the criterion for ferromagnetic state we differentiate Eq. (\ref{28}) with respect to $m$. As a result, we get two terms. The first one is the change of the DOS with rising 
$m$. It depends on the self-energies $\Sigma _{1,2}^\sigma$ and $\Sigma _U^\sigma$ and contributes to the correlation factor $K$, see Eq. (\ref{32}) below. The second term is 
the spin-dependent band shift, which contributes merely to the Hartree field. In effect we obtain the following formula
\be
1 = K + \left[ {I^{cr}  + zJ + z\Delta t(1 - n)} \right]\int\limits_{ - \infty }^\infty  {\rho _{m = 0}^\sigma  \left( \varepsilon  \right)P_T \left( \varepsilon  \right)d\varepsilon } 
\label{31}
\ee
where
\be
\rho ^\sigma  (\varepsilon ) =  - \frac{1}{\pi }{\mathop{\rm Im}\nolimits} F^\sigma  (\varepsilon ),
\;\;\;\;P_T \left( \varepsilon  \right) = f^2 \left( \varepsilon  \right)e^{\beta \left( {\varepsilon  - \mu } \right)} \beta,
\;\;\;\;K =  - \frac{2}{\pi }{\mathop{\rm Im}\nolimits} \int\limits_{ - \infty }^\infty  {f\left( \varepsilon  \right)\frac{{\partial F^\sigma  \left( \varepsilon  \right)}}{{\partial 
m}}d\varepsilon}
\label{32}
\ee
and $I^{cr}$  is given by Eq. (\ref{5}). In the zero temperature limit the function $P_T \left( \varepsilon  \right)$  becomes the Dirac's delta function and we have
\be
1 = K + \rho \left( {\varepsilon _F } \right)\left[ {I^{cr}  + zJ + z\Delta t\left( {1 - n} \right)} \right],
\label{33}
\ee
where
\be
\rho (\varepsilon _F ) =  - \frac{1}{\pi }{\mathop{\rm Im}\nolimits} \left. {F^\sigma  (\varepsilon _F )} \right|_{m = 0},
\;\;\;\;K =  - \frac{2}{\pi }{\mathop{\rm Im}\nolimits} \int\limits_{ - D}^{\varepsilon _F } {\frac{{\partial F^\sigma  \left( \varepsilon  \right)}}{{\partial m}}d\varepsilon }
\label{34}
\ee
and $\varepsilon _F$  is the Fermi energy calculated from Eq. (\ref{28}) in the zero temperature limit. 

Using relation
\be
\frac{{\partial F^\sigma  }}{{\partial m}} = \frac{{\partial F^{\sigma}  }}{{\partial \Sigma _{U}^{\sigma} }}\frac{{\partial \Sigma _U^\sigma  }}{{\partial m}} + \frac{{\partial 
F^\sigma  }}{{\partial \Sigma _{1,2}^\sigma  }}\frac{{\partial \Sigma _{1,2}^\sigma  }}{{\partial m}} + \frac{{\partial F^\sigma  }}{{\partial \alpha ^\sigma  }}\frac{{\partial \alpha 
^\sigma  }}{{\partial m}},
\label{35}
\ee
where $\alpha ^\sigma   = 2\frac{{\Delta t}}{t}n^{ - \sigma }$, we obtain the correlation factor as the sum of the on-site, inter-site and assisted hopping correlation factors
\be
K = K_U  + K_{ij}  + K_{\Delta t}, 
\label{36}
\ee
where
\[
K_U  =  - \frac{2}{\pi }{\mathop{\rm Im}\nolimits} \int\limits_{ - D}^{\varepsilon _F } {\frac{{\partial F^\sigma  }}{{\partial \Sigma _U^\sigma  }}\frac{{\partial \Sigma _U^\sigma  
}}{{\partial m}}d\varepsilon } ,\quad \quad K_{ij}  =  - \frac{2}{\pi }{\mathop{\rm Im}\nolimits} \int\limits_{ - D}^{\varepsilon _F } {\frac{{\partial F^\sigma  }}{{\partial \Sigma 
_{1,2}^\sigma  }}\frac{{\partial \Sigma _{1,2}^\sigma  }}{{\partial m}}d\varepsilon },
\]
\be
 K_{\Delta t}  =  - \frac{2}{\pi }{\mathop{\rm Im}\nolimits} \int\limits_{ - D}^{\varepsilon _F } {\frac{{\partial F^\sigma  }}{{\partial \alpha ^\sigma  }}\frac{{\partial \alpha 
^\sigma  }}{{\partial m}}d\varepsilon }.
\label{37}
\ee
 To establish the link with the existing literature, it can be written on the basis of Ref. \cite{23} for the single-site correlation factor in the case of strong correlation $\left( {U\gg D} 
\right)$ that
\be
K_U  =  - \frac{1}{\pi }{\mathop{\rm Im}\nolimits} \ln \left( {\frac{{\Sigma _U^\sigma  }}{{U - \Sigma _U^\sigma  }}} \right).
\label{38}
\ee
For the two-site correlation factor, using Eq. (\ref{30}) in Eq. (\ref{37}), we obtain
\be
K_{ij}  =  - \frac{2}{\pi }{\mathop{\rm Im}\nolimits} \int\limits_{ - D}^{\varepsilon _F } {\left\{ {\frac{1}{{b^\sigma  }}\frac{{\partial \Sigma _{1,2}^\sigma  \left( \varepsilon  
\right)}}{{\partial m}}\left[ {\frac{{F^\sigma  }}{t} + \frac{{\partial F_0^\sigma  }}{{\partial \Sigma _{1,2}^\sigma  }}} \right]} \right\}d\varepsilon },
\label{39}
\ee
where $$F_0^\sigma   = F_0 \left( {\frac{{\varepsilon  - \Sigma _U^\sigma  }}{{b^\sigma  }}} \right).$$

Later on we used the explicit form of $F^\sigma  \left( {\varepsilon ,m} \right)$, obtained after the elimination of self-energy $\Sigma _U^\sigma$. Although this elimination is only 
possible in some specific cases, e.g. the semi-elliptic DOS, there is a great advantage of the simplicity in this case since there is no need to develop a formula for the correlation 
factors any further than has been formulated in Eq. (\ref{34}).

For the assisted hopping correlation factor, using Eq. (\ref{30}) in Eq. (\ref{37}), we obtain
\be
K_{\Delta t}  = \frac{2}{\pi }\frac{{\Delta t}}{t}{\mathop{\rm Im}\nolimits} \int\limits_{ - D}^{\varepsilon _F } {\left\{ {\frac{1}{{b^\sigma  }}\left[ {F^\sigma  \left( \varepsilon  
\right) + \frac{{\partial F_0^\sigma  }}{{\partial \alpha ^\sigma  }}} \right]} \right\}d\varepsilon }, 
\label{40}
\ee
where $F_0^\sigma$  was defined earlier after Eq. (\ref{39}).

In the lowest order Hartree-Fock approximation; $\Sigma _{1,2}^\sigma   \approx \bar \varepsilon _{1,2}^\sigma$, $\Sigma _U^\sigma   = Un^{ - \sigma }$,  $\Sigma _U^\sigma   = 
0$ and from Eq. (\ref{33}) with the help of Eqs (\ref{39}) and (\ref{40}) we obtain
\be
1 = \rho \left( {\varepsilon _F } \right)\left[ {I^{cr}  + zJ + U + \frac{{2\varepsilon _F }}{t}\left( {\frac{{\partial \bar \varepsilon _{1,2}^\sigma  }}{{\partial m}} - \Delta t} \right)} 
\right].
\label{41}
\ee
 The factor  $I^{cr}  + zJ + U$ above comes from the Hartree-Fock approximation, the extra term in the curl bracket; $\left( {2\varepsilon _F /t} \right)\left( {\left( {\partial \bar 
\varepsilon _{1,2}^\sigma  /\partial m} \right) - \Delta t} \right)$, comes from the inter-site and assisted hopping correlation factors. These factors collect contributions to the 
magnetic criterion coming from the change of DOS over the whole energy interval, as opposed to the contribution coming only at the Fermi energy, as in the case of the classic 
Hartree-Fock approximation (the first Stoner type term in Eq. (\ref{41})).

It is worthwhile to note here that in the first order approximation; $\Sigma _U^\sigma   \approx Un^{ - \sigma }$  and the on-site correlation factor $K_U$  is equal to zero. Without 
this on-site correlation factor $\left( {K_U  = 0} \right)$, together with $\Sigma _{1,2}^\sigma   \approx \bar \varepsilon _{1,2}^\sigma$, as well as $\Delta t = 0$, we obtained from 
Eq. (\ref{41}), for the constant DOS, the Hirsch's result (see Ref. \cite{15}) 
\be
j = \frac{{2\left[ {1 - {{(I_{in}^{cr}  + U)} \mathord{\left/
 {\vphantom {{(I_{in}^{cr}  + U)} D}} \right.
 \kern-\nulldelimiterspace} D}} \right]}}{{1 + a - \left( {3a - 1} \right)\left( {1 - n} \right)^2 }},
\label{42}
\ee
 where $aj = \frac{{J - V}}{{2t}} \equiv \frac{{\varepsilon _1 }}{{2t}}$ and $j = \frac{{J + J'}}{{2t}} \equiv \frac{{\varepsilon _2 }}{{2t}}.$
 
 From Eq. (\ref{41}), without the correlation factors, the Stoner criterion of magnetism was reached in the following well-known form 
 \be
1 = \rho ^0 \left( {\varepsilon _F } \right)\left( {I^{cr}  + U} \right)
 \label{43}
 \ee
 where $\rho ^0 \left( {\varepsilon _F } \right)$  is the unperturbed DOS on the Fermi level.
 
Numerical results in the next section were calculated on the basis of equation (\ref{33}), which was obtained in the zero temperature limit of Eq. (\ref{31}). The correlation factors 
in this limit are defined by Eqs (\ref{37}). 

\vskip0.5cm 

\noindent {\Large {\bf 4. Numerical examples}} 
\vskip0.5cm 

This new model will be illustrated by showing the dependence of the critical on-site field versus the carrier concentration; $I^{cr}(n)$. It must be remembered that the relation 
between this field and the total molecular field $M$ computed here is as follows $I=M-z \Delta t(1-n)-zJ$. Only the first order approximation was considered for the two-site 
self-energy $\Sigma _{1,2}^\sigma   \cong \bar \varepsilon _{1,2}^\sigma$. The single-site self-energy $\Sigma _U^\sigma$ was solved using the standard CPA equation (see Eq. 
(\ref{25})). The perturbed Slater-Koster function has the form
\be
F^\sigma  \left( \varepsilon  \right) = \frac{1}{{b^\sigma  }}F_0 \left( {\frac{{\varepsilon  - \Sigma _U^\sigma  }}{{b^\sigma  }}} \right),
\label{44}
\ee
where $F_0(\varepsilon)$ is the unperturbed Slater-Koster function, for which we assume the semi-elliptic form 
\be
F_0 \left( \varepsilon  \right) = \frac{2}{D}\left[ {\frac{\varepsilon }{D} - \sqrt {\left( {\frac{\varepsilon }{D}} \right)^2  - 1} } \right]
\label{45}
\ee
and $b^\sigma   = 1 - \frac{{\bar \varepsilon _{1,2}^\sigma  }}{t} - 2\frac{{\Delta t}}{t}n^{ - \sigma }$.

To illustrate the new results we used Eq. (\ref{33}) to calculate the dependence of critical (minimal) value of $I^{cr}$ on the carrier concentration $n$.  The inter-site correlation 
factors $K_{ij}$ and $K_{\Delta t}$ are calculated from Eq. (\ref{39}) and Eq. (\ref{40}) with the help of function $F^{\sigma}$ defined by Eq. (\ref{44}). Inter-site interactions 
$\varepsilon_1$  and  $\varepsilon_2$, parameter $p$ and the asymmetry parameter $a$ are defined as follows 
\be
\varepsilon _1  \equiv (J - V) = ap\frac{D}{z} = apt
{\;\;\;\;\rm and \;\;\;\;}
\varepsilon _2  \equiv J + J' = p\frac{D}{z} = pt.
\label{46}
\ee
We assume that the inter-site exchange and pair hopping interactions are equal $J=J'$  (see Ref. \cite{19}), then $J=pD/2z=pt/2$.

In all the figures presented below, the dependence $I^{cr}(n)$ is expressed in units of half-bandwidth $D$. 

\begin{figure}[t]
\begin{center}
\epsfig{file=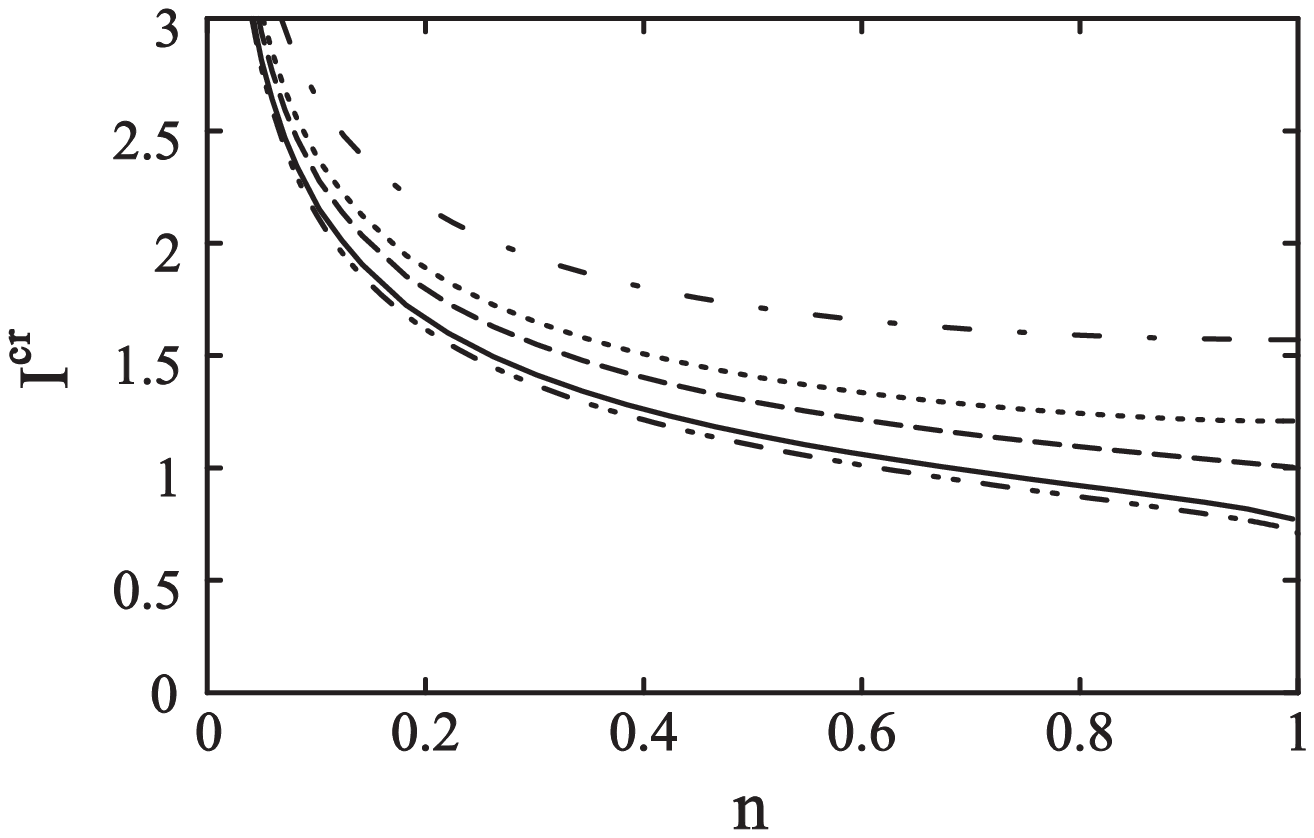,width=0.45\hsize}
 \par\vspace{1.5ex}\makebox[0.5\hsize]
    {\small FIG. 1}
	\end{center}
	\vskip0.5 cm
	\caption{Dependence of critical field on carrier concentration $I^{cr}(n)$, for the single-site correlation alone $(U\neq 0, J=J'=V=\Delta t=0)$, for different strengths of $U$; 
$U=5D$ is the solid line, $U=D$ the dashed line, $U=0.5D$ the dotted line. For comparison we show the $I^{cr}(n)$ dependence without on-site Coulomb correlation - dot-dashed 
line and for the strong single-site correlation; $U/D=\infty$ is the double dot-dashed line. }
\vskip0.5 cm
\end{figure}

The function $I^{cr}(n)$ for the single-site correlation $(U\ne 0)$ alone $(J=J'=V=\Delta t=0)$ is shown in Fig. 1. The solid line corresponds to $U=5D$, the dashed line is for 
$U=D$, and the dotted line is for $U=0.5D$. For comparison, it is presented in the same figure the dependence of $I^{cr}(n)$ without on-site Coulomb correlation (the dot-dashed 
line for $K_U=0$, which is the standard Stoner criterion for magnetism) and for the infinitely strong single-site correlation; $U/D = \infty$  (the double dot-dashed line). 

Analyzing the $I^{cr}(n)$ curves presented in Fig. 1 one can see that the on-site correlation $K_U$ decreases the minimal value of atomic field $I^{cr}$ necessary to obtain the 
ferromagnetic state. The dominant reduction takes place at the half-filling, already at relatively small $U \approx 3D$. Further increase of $U$ does not influence the ferromagnetic 
state much.

In Figs 2-4, we have presented the dependence of critical on-site interaction $I^{cr}(n)$ in the presence of inter-site correlations. Figs 2 and 3 do not have hopping interaction 
$(\Delta t=0)$. With the symmetric initial DOS (semi-elliptic), the curves for these interactions remain symmetric with respect to $n=1$. This is why the results in Figs 2 and 3 are 
drawn only for $0 \le n \le 1$.

\begin{figure}[t]
\begin{center}
\epsfig{file=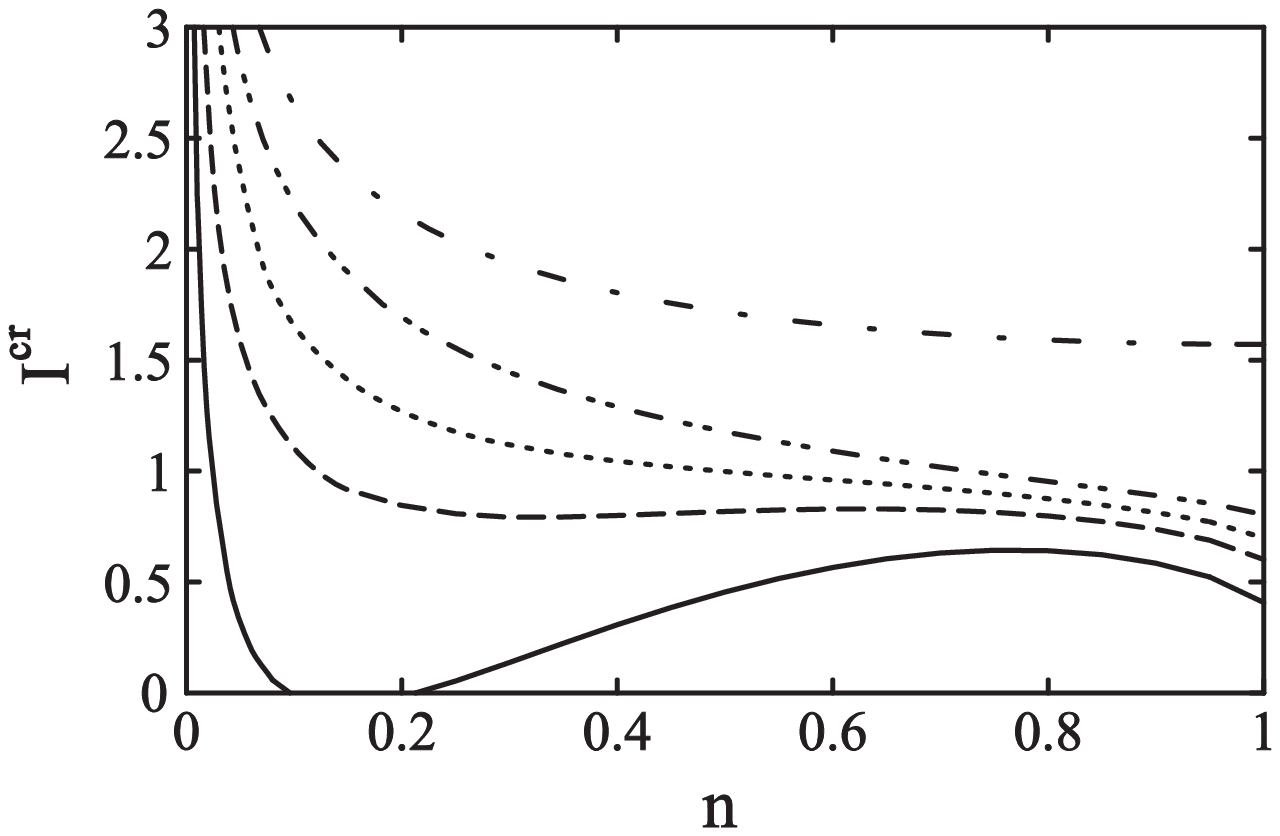,width=0.45\hsize}
 \par\vspace{1.5ex}\makebox[0.5\hsize]
    {\small FIG. 2}
	\end{center}
	\vskip0.5 cm
	\caption{Dependence of critical field on carrier concentration $I^{cr}(n)$, at $U=3D$, $a=-1.5$ and $\Delta t=0$ for different strengths of inter-site correlation (different 
parameter $p$); $p=1$ is the solid line, $p=0.5$ the dashed line, $p=0.25$ the dotted line. For comparison we show the $I^{cr}(n)$ dependence for $p=0$ the double dot-dashed 
line, and for both inter-site correlation $p=0$ and on-site correlation $K_U=0$ is the dot-dashed line. }
\vskip0.5 cm
\end{figure}

Fig. 2 shows the influence of parameter $p=z(J+J')/D=(J+J')/t$ on the value of on-site critical field $I^{cr}$. The solid line is for $p=1$, the dashed line for $p=0.5$, and the dotted 
line for $p=0.25$. The Coulomb interaction; $U=3D$ and the asymmetry parameter of inter-site interactions; $a = \frac{{\varepsilon _1 }}{{\varepsilon _2 }} = \frac{{J - V}}{{J + J' 
}} =  - 1.5$. We have shown in the same figure for comparison the dependence of $I^{cr}(n)$ for the standard Stoner criterion of ferromagnetism (the dot-dashed line) and 
$I^{cr}(n)$ for single-site correlation only; $U=3D$ and $p=0$ (double-dot-dashed line). Analyzing the presented curves, it is possible to see that the inter-site correlations 
significantly decrease the value of the intra-site critical field necessary for magnetism $I^{cr}$. Equivalently the decrease of parameter $p$ increases the minimal value of atomic 
field $I^{cr}$. The decrease of $p$ means the decrease of inter-site interactions which causes the decrease of inter-site correlation factor $K_{ij}$ and the increase of DOS on the 
Fermi level $\rho \left( {\varepsilon _F } \right)$ especially for $n$ close to zero or two.

\begin{figure}[t]
\begin{center}
\epsfig{file=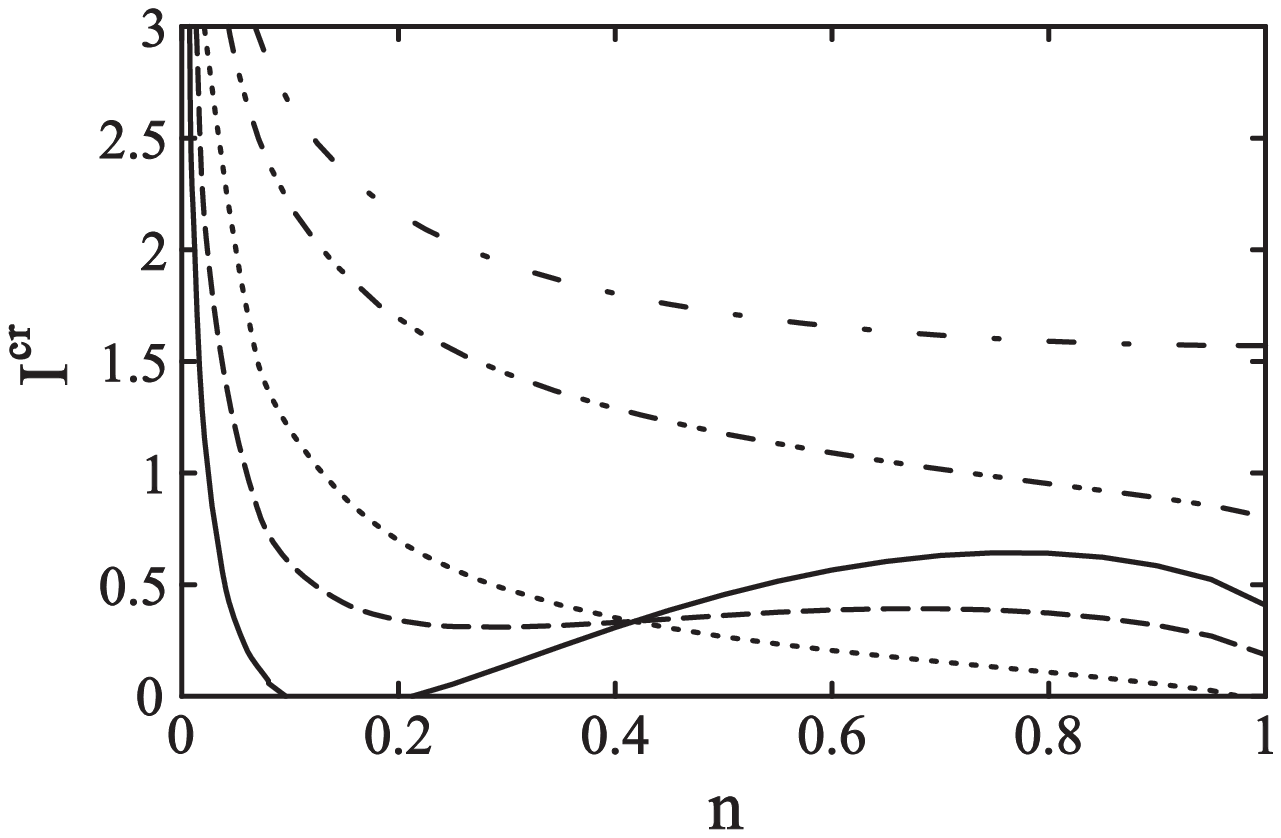,width=0.45\hsize}
 \par\vspace{1.5ex}\makebox[0.5\hsize]
    {\small FIG. 3}
	\end{center}
	\vskip0.5 cm
	\caption{Dependence of critical field on carrier concentration $I^{cr}(n)$, at $U=3D$, $p=1$ and $\Delta t=0$ for different asymmetry parameters $a$; $a=0.5$ is the dotted line; 
$a=-0.5$ the dashed line; $a=-1.5$ the solid line. For comparison we show the $I^{cr}(n)$ dependence for $p=0$ is the double dot-dashed line, and for both inter-site correlation 
$p=0$ and on-site correlation $K_U=0$ is the dot-dashed line.}
\vskip0.5 cm
\end{figure}

The role of factor $a$ in minimising the critical field for magnetism is shown in Fig 3. The solid line is for $a=-1.5$, dashed line for $a=-0.5$ and the dotted line for $a=0.5$. All the 
curves have $p=1$ and $U=3D$. We have shown in the same figure for comparison the dependence of $I^{cr}(n)$ for the standard Stoner criterion of ferromagnetism (the 
dot-dashed line), and $I^{cr}(n)$ for the single site-correlation only; $U=3D$ and $p=0$ (the double-dot-dashed line). Analyzing $I^{cr}(n)$ curves for different parameters $a$ 
one can see that the decrease of $a$ causes the increase of $I^{cr}$ for concentrations close to half-filling but it decreases $I^{cr}$ at small concentrations and concentrations 
close to full-filling. This effect is the result of change in two factors; inter-site correlation factor $K_{ij}$ and the DOS on the Fermi level $\rho \left( {\varepsilon _F } \right)$. The 
value of $\rho \left( {\varepsilon _F } \right)$ increases with $a$, but the value of inter-site correlation factor $K_{ij}$ decreases with increasing $a$. The sign of parameter $a$  
depends on the difference between interactions $J$ and $V$. For typical 3-d metal we have $J<V$ \cite{1,21} and parameter $a$ is negative. However, when $J>V$ \cite{15} or 
$V<0$  (which can take place in superconducting cuprates) the sign of $a$ will be positive. 

As can be seen from Figs 2 and 3, at some values of parameters $p$ and $a$, we have obtained the ferromagnetic state already at zero value of intra-site field. For the optimal 
values of $p=1$ and $a=-1.5$ the ferromagnetic state exists at concentrations $n=0.09 - 1.91$ and also at $n=1.78 - 1.91$, the latter is more interesting from the point of view of 
ferromagnetic elements.

\begin{figure}[t]
\begin{center}
\epsfig{file=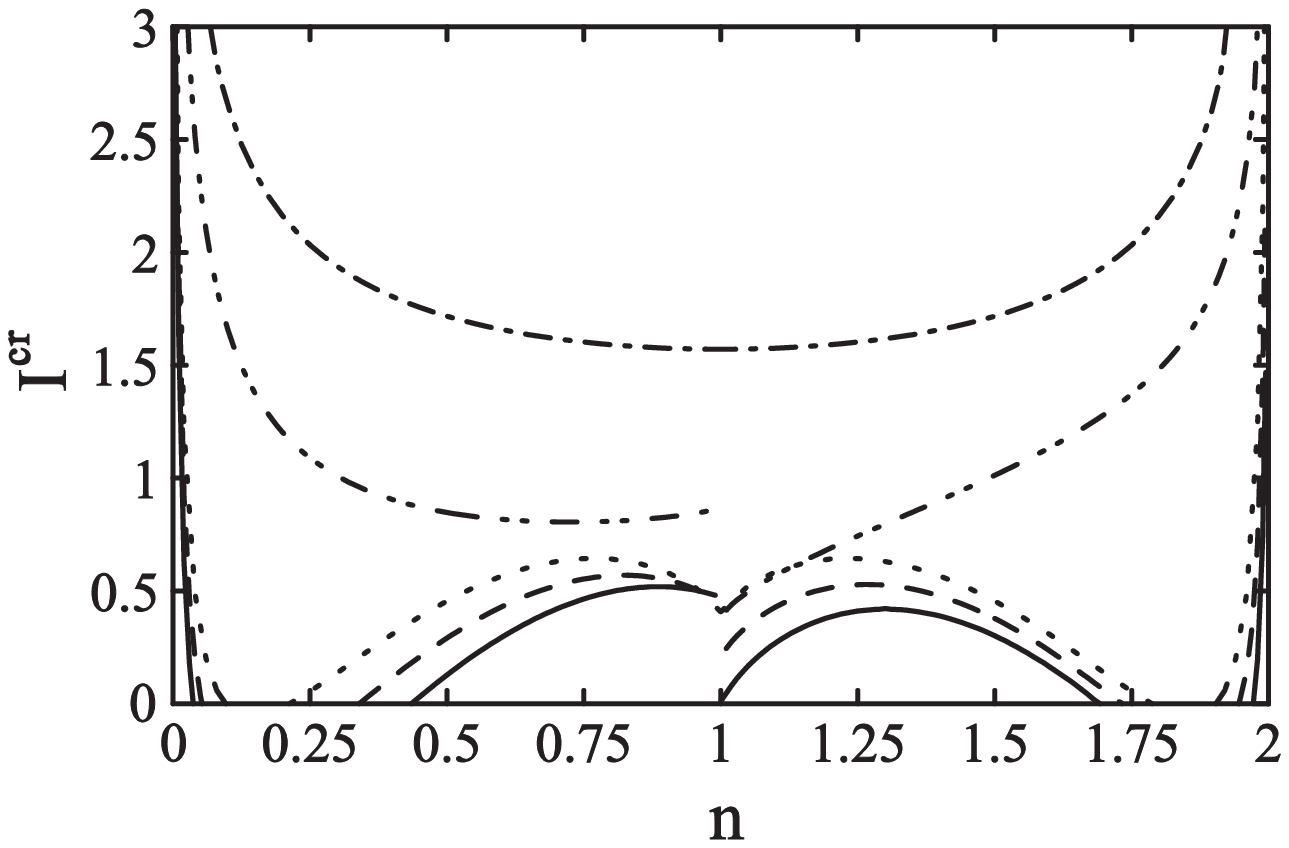,width=0.45\hsize}
 \par\vspace{1.5ex}\makebox[0.5\hsize]
    {\small FIG. 4}
	\end{center}
	\vskip0.5 cm
	\caption{Dependence of the critical field on carrier concentration $I^{cr}(n)$, for $a=-1.5$ and $p=1$ in the case of correlation $U=3D$ for different values of parameter $\Delta 
t$; $\Delta t=0.2t$ is the solid line, $\Delta t=0.1t$ the dashed line, $\Delta t=0$ the dotted line. For comparison we show the $I^{cr}(n)$  dependence for $p=0$, $\Delta t=0.2t$ is 
the double dot-dashed line, and for both inter-site correlation $p=0$ and on-site correlation $K_U=0$ is the dot-dashed line.}
\vskip0.5 cm
\end{figure}

In Fig. 4 we show the critical value of the internal exchange field $I^{cr}$ in function of carrier concentration at $a=-1.5$, $p=1$ and for different values of the hopping interaction 
$\Delta t$ in the case of single-site Coulomb correlation $U=3D$; $\Delta t=0$ is the solid line, $\Delta t=0.1t$ the dashed line, and $\Delta t=0.2t$ the dotted line. We show in the 
same figure for comparison the dependence $I^{cr}(n)$ for $p=0$,$\Delta t=0.2t$ is the double dot-dashed line and without both inter-site and on-site correlation $(K=0)$ the  
dot-dashed curve. The increase of the hopping interaction $\Delta t$ causes for all electron occupations $n$ to decrease in minimal value of the internal exchange field creating 
ferromagnetism $I^{cr}$. However, a closer look into Fig. 4 reveals that the largest decrease takes place for carrier concentrations below half-filling. As a result, the hopping 
interaction helps ferromagnetism more in the beginning of the 3-d row of elements than at the end. The increase of the ratio; $I/D$ along the 3-d row, as well as the increase of 
other interaction constants along this row (see Ref. \cite{26}), can offset this effect. At half filling, we have discontinuity of $I^{cr}(n)$. Coming from the left and right side with 
$n$ one has 
\[
I^{cr}  \approx \frac{{1 - K_U  - K_{ij} }}{{\rho \left( {\varepsilon _F } \right)}}\left( {1 \pm 2\frac{{\Delta t}}{t}} \right) - zJ - z\Delta t(1 - n),
\]
with the upper and lower sign respectively. The discontinuity exists only in the split band limit, which takes place for our $U$ (our $U=3D$). It is absent in the paper \cite{14}, 
since the Coulomb correlation was treated there in the Hartree-Fock approximation which can only shift the spin bands and is unable to change their shapes and to split them into 
sub-bands when $U$ is high enough.

\vskip0.5cm 
\newpage
\noindent {\Large {\bf 5. Conclusions}} 
\vskip0.5cm 

As can be seen from Fig. 1 the single-site electron correlation, $K_U$, seriously helps ferromagnetism, especially for concentrations of $n$ close to half-filling. For the 3-d group 
of elements, the enhancement will not be the same for small and large $n$ as we have to take into account the fact that the ratio of $I/D$ existing in different elements growths with 
$n$. This increase will increase the chances of ferromagnetism towards the end of the 3d-group of elements.

As was already stated, the strong single-site Coulomb correlation $U>D$ favours the ferromagnetism (but rather antiferromagnetism) at half-filled band (see Fig. 1). Comparing this 
result with the 3-d group of elements, it is possible that this situation would correspond to antiferromagnetic elements like Cr, and Mn, which have concentrations of electrons 
close to half-filling. The cohesion energy,$E_{\rm coh}$, strongly decreases for these elements which gives evidence of the decreasing bandwidth $2D$ to a small value. Such a 
decrease of the bandwidth follows from our Eq. (\ref{27}) with increasing interaction constants. In the first order approximation it can be written as; $D_{\rm eff}=D[1-p(1+a)I_0]$, 
where $I_0$ has maximum at half-filling. The small value of effective bandwidth would justify using models with $U\gg D$ for these two elements.

In extending description of ferromagnetism by the CPA method we included;
\begin{itemize}
\item [(i)]{spin dependent band shift coming from the intra- and inter-site exchange interactions ($I$ and $J$, $V$) and assisted hopping interaction $\Delta t$,}
\item [(ii)]{spin dependent change of the band width depending on electron concentrations and all the interactions. Narrowing of the band increases DOS on the Fermi level $\rho 
\left( {\varepsilon _F } \right)$,}
\item [(iii)]{band shape changes through the on-site $(K_U)$ and inter-site correlation factors ($K_{ij}$ and $K_{\Delta t}$).}
\end{itemize}

Figs 2-4 show the role of inter-site correlations for magnetism. Comparing the curves with different parameter $p=z(J+J')/D$ in Fig. 2 one can see that the inter-site correlation 
additionally decreases the intra-atomic field necessary to create the ferromagnetism. The value of on-site Coulomb repulsion was fixed at the realistic level of $U=3D$. The effect is 
particularly strong for large but still realistic $p=1$. One can see that the decrease of intra-atomic field (symmetrical in $n=1$) is even larger than the one coming from the 
single-site correlation $K_U$ and it takes the place for values of $n$ corresponding better to ferromagnetic elements. The influence of different inter-site interactions on 
ferromagnetism is given by the parameter $a = \frac{{\varepsilon _1 }}{{\varepsilon _2 }} = \frac{{J - V}}{{J + J'}}$. 

\noindent
It is strongest at both ends of the band and for $a$ as small as possible, we assume $a=-1.5$, see Fig 3. Such a value is possible for transition metals, since for these elements 
values of inter-site charge-charge interaction $V$ are positive and larger than the inter-site exchange interaction $J$. For $a=1$, i.e. $J+J'=J-V$ the inter-site interactions do not 
change the effective bandwidth (see Eq. (\ref{21})), hence the inter-site correlation factor $K_{ij}=0$ and the only influence of inter-site interactions on ferromagnetism is by 
increasing the total molecular field $M$.

To sum up, we can say that the inter-site correlation in co-operation with single-site correlation enhances ferromagnetism preferably at the end of the band, or speaking about 3-d 
row, at the end of this row.

Fig. 4 shows that the hopping interaction $\Delta t$ also enhances magnetism, but this effect is stronger for smaller concentrations of $n$ than for larger concentrations. This 
result seems to be in contradiction to the experimental evidence on the 3-d row of elements, but the increase of the ratio $I/D$ and of other interactions along the 3-d row of 
elements can offset this effect. 

In general the CPA method used for the Hubbard model, without any additional field of on-site or inter-site origin, does not yield the ferromagnetic state. It comes out of the fact 
that the spin densities are not shifted or twisted with respect to each other $\rho _ \uparrow  \left( \varepsilon  \right) = \rho _ \downarrow  \left( \varepsilon  \right)$. For 
comparison, the Spectral Density Approach (SDA) \cite{6}, already mentioned above, gives the ferromagnetic state for Hubbard model at $U>D$. The main advantage of the SDA 
is obtaining spin band shifts due to 'higher hopping $t_{ij}$ correlations'. The drawback of this method is that the self-energy is real which neglects quasiparticle damping, that 
takes place in these materials \cite{7}. Comparing the SDA method with our model, one notices that the so called "higher" correlation function (defined as $B_{-\sigma}$ in 
\cite{7}) is the interaction very similar to the assisted hopping interaction $\Delta t$ used in our paper. Using moments method, introduced by Harris and Lange \cite{27} and more 
recently by Herrmann and Nolting \cite{7}, the CPA technique was applied to two modified atomic levels with modified probabilities, by the spin depended shifted atomic level 
$B_{-\sigma}$. This approach was called Modified Alloy Analogy (MAA). As a result, Herrmann and Nolting obtained within MAA, the ferromagnetic state for some 
concentrations with the maximum of magnetic moment around $n\approx 0.7$. Using the SDA method, the same authors obtained ferromagnetism for concentrations $n>0.55$. 
Using our model, with only the assisted hopping interaction $\Delta t$, we obtained very small values of the on-site exchange interaction $I$ necessary for ferromagnetism at 
$n\approx 0.7$ (see double dot-dashed line in Fig. 4). This result is very similar to the MAA result mentioned above.  Introducing into consideration the other inter-site 
interactions, we got ferromagnetism for smaller concentrations and even at zero values of the on-site exchange interaction $I$ (see Fig.4).

The ferromagnetic state in the standard Hubbard model can be obtained for the infinite coordination number $(Z\rightarrow \infty)$ using dynamical mean-field theory (DMFT). 
This method, if used for symmetrical DOS (like the one used by us), needs very large values of on-site Coulomb correlation $U$. For intermediate $U$ in DMFT method to obtain 
the ferromagnetic state, one has to use strongly asymmetric DOS (see e.g. Ref. \cite{10}). The use of such DOS in CPA would also allow for a large reduction of the inter-atomic 
and intra-atomic field necessary to create the ferromagnetism. Vollhardt and co-workers \cite{10,28} pointed out that use of the inter-site exchange and other nearest-neighbour 
interactions, which was done in our paper using a different method, would reduce the value of $U$ necessary for ferromagnetism in DMFT method.

In general the Hartree-Fock approximation overestimates the ordering. As a result, the same interactions when treated in higher order approximation, required much larger 
interaction constants to obtain the alignment. On the other hand, use of the Hartree-Fock approximation in this paper may be justified since, for many electron occupations, the 
magnitude of the field used, is very small as compared to the kinetic energy characterized by the bandwidth.

\end{document}